\documentstyle[12pt]{article}
  \newcounter{c_one}
  \newcommand{\be}{\begin{equation}}
  \newcommand{\ee}{\end{equation}}
  \newcommand{\bea}{\begin{eqnarray}}
  \newcommand{\eea}{\end{eqnarray}}
  \newcommand{\lb}{\label}
\renewcommand{\a}{\alpha}

\renewcommand{\d}{\delta}
  
\renewcommand{\l}{\lambda}
\renewcommand{\L}{\Lambda}
  
\renewcommand{\S}{\Sigma}

  \newcommand{\bcal}[1]{{\cal #1}}
  \newcommand{\cala}{\bcal{A}}
  \newcommand{\calb}{\bcal{B}}
  \newcommand{\ad}{\mbox{\rm ad}}
  \newcommand{\Ad}{\mbox{\rm Ad}}
\renewcommand{\Re}{\,\mbox{\rm Re}\,}
\renewcommand{\Im}{\mbox{\rm Im}\;}
  \newcommand{\ldel}{\langle}
  \newcommand{\rdel}{\rangle}
  \newcommand{\sprod}    [2]{\ldel \,#1\,,\,#2\,\rdel}
  \newcommand{\Bigsprod} [2]{\Bigl \langle\,#1\,,\,#2\,\Bigr \rangle}

  \newcommand{\lbrac}[2]{[#1,#2]}
  \newcommand{\lbrak}[2]{[#1,#2]}
  \newcommand{\Min}{\mathop{\rm Min}}
  \newcommand{\supp}{\mathop{\rm supp}}
  \newcommand{\bra}[1]{\ldel\,#1\,|}
  \newcommand{\ket}[1]{|\,#1\,\rdel}
  \newcommand{\braket}[2]{\ldel\,#1\,|\,#2\,\rdel}
  
  \newcommand{\bmrm}[1]{\mbox{\rm #1}}

\begin{document}

\begin{titlepage}
\begin{flushright}
ZU-TH 19/94
\end{flushright}
\vspace{3 ex}
\begin{center}
\vfill
{\LARGE\bf Instability Proof for Einstein-Yang-Mills Solitons with
arbitrary Gauge Groups}
\vfill
{\bf Othmar Brodbeck and Norbert Straumann}
\vskip 0.5cm
Institute for Theoretical Physics\\University of Z\"urich\\
Winterthurerstrasse 190, CH-8057 Z\"urich
\end{center}
\vfill
\begin{quote}
We prove that static, spherically symmetric, asymptotically flat,
regular solutions of the Einstein-Yang-Mills equations are unstable for
arbitrary gauge groups, at least for the ``generic" case. This
conclusion is derived without explicit knowledge of the possible
equilibrium solutions.
\end{quote}
\vfill
\end{titlepage}
\section{Introduction}
In several recent papers \cite{NS1,OB1,OB1.1,OB2} we have studied
important aspects of the Einstein-Yang-Mills (EYM) system for arbitrary
gauge groups. In particular, we investigated the classification and
properties of spherically symmetric EYM solitons (magnetic structure,
Chern-Simons numbers) and a generalization of the Birkhoff theorem for
the non-Abelian case. We also worked out the generalization of the
first law of black hole physics (Bardeen-Carter-Hawking formula),
allowing for additional Higgs and dilaton fields
\cite{heusler1,heusler2}. For other studies of these and related topics
we refer to \cite{sudarsky,okai,rogatko}.

In the present paper we prove that static, spherically symmetric,
asymptotically flat, regular solutions of the EYM equations for any
gauge group are unstable, at least in the ``generic" case (defined
below). In a recent letter \cite{OB3} we have already sketched how we
arrived at this result. Here, we present full details of the proof and
discuss also some further mathematical issues involved.

On physical grounds, this instability was expected, because we had
shown earlier that the Bartnik-McKinnon solutions \cite{bartnik} for
the gauge group SU$(2)$, as well as the related black hole solutions
\cite{volkov,bizon,kunzle1} are unstable \cite{NS2,NS3,zhou1,zhou2}. A
mathematical proof of this expectation presents, however, quite a
challenge, since one can not rely on any knowledge of the possible
solutions (apart from regularity and boundary conditions).

Our strategy is based on the study of the pulsation equations
describing linear radial perturbations of the equilibrium solutions and
involves the following main steps: First, we show that the frequency
spectrum of a class of radial perturbations is determined by a coupled
system of radial ``Schr\"odinger equations". Eigenstates with negative
eigenvalues correspond to exponentially growing modes. Using the
variational principle for the ground state it is then proven that there
always exist unstable modes (at least for ``generic" solitons).

There is, unfortunately, no direct way to apply our method to black
holes, because of problems related to the boundary conditions at the
horizon. We have, however, recently used a similar procedure to
establish also the instability of gravitating regular sphaleron
solutions of the SU$(2)$ Einstein-Yang-Mills-Higgs system with a
SU$(2)$ Higgs doublet \cite{boschung}, which have been constructed
numerically in \cite{greene}.

The paper is organized as follows: In Sec.\ 2 we recall some basic
facts and equations of our previous work \cite{OB1,OB2} which will be
needed in the present analysis. In Sec.\ 3 we then derive the
linearized perturbation equations for solitons (and black holes) and
bring them into a convenient partially decoupled form. The resulting
eigenvalue problem is discussed in Sec.\ 4, and in Sec.\ 5 we show the
existence of unstable modes.
\section{Spherically symmetric EYM fields}
We begin with a convenient description of gauge fields with spherical
symmetry (for derivations see \cite{OB1}). Let us fix a maximal torus
$T$ of the gauge group $G$ with the corresponding integral lattice $I$
( = kernel of the exponential map restricted to the Lie algebra $LT$ of
the torus $T$). In addition, we choose a basis $S$ of the root system
$R$ of real roots. The corresponding fundamental Weyl chamber
\be
K(S)=\{\, H\in LT \mid \alpha (H)>0 \,\mbox{ \rm for all } \alpha\in S
\,\} \lb{ns1}
\ee
plays an important role in what follows. We have shown in \cite{OB1}
that to a given spherically symmetric gauge field configuration, there
belongs in a natural way a canonical element $H_\lambda\in
I\cap\overline{K(S)}$ which characterizes the corresponding principal
bundle $P(M,G)$ over the spacetime manifold $M$ admitting an SU$(2)$
action. If the solution is regular at the origin,  $H_\lambda$ lies in
a small finite subset of $I\cap\overline{K(S)}$, which we have
described in \cite{OB2}. In much of our discussion we exclude (for
technical reasons) the possibility that $H_\lambda$ lies on the
boundary of the fundamental Weyl chamber. The term {\em generic} always
refers to fields for which the classifying element $H_\lambda$ is
contained in the {\em open} Weyl chamber $K(S)$.

The SU$(2)$ action on $P(M,G)$ by bundle automorphisms induces an
action on the base manifold $M$. An SU$(2)$-invariant connection in
$P(M,G)$ defines a connection in each subbundle over a single orbit of
this induced action, which by Wangs theorem is described by a linear
map $\L\colon L\bmrm{SU}(2)\rightarrow LG$, depending smoothly on the
orbit and satisfying
\be
\L_1=[\L_2,\L_3]\:, \qquad
\L_2=[\L_3,\L_1]\:, \qquad
\L_3 =-H_\l/4\pi\:.
\lb{ns2}
\ee

Here, $\L_k:=\L(\tau_k)$ and $2i\tau_k$ are the Pauli matrices. These
equations imply that $\L_+:=\L_1+i\L_2$ lies in the following direct
sum of root spaces $L_\alpha$ of $LG_\bcal{C}$:
\be
\renewcommand{\arraystretch}{1.3}
\begin{array}{cc}
{\displaystyle
\Lambda_+ \in  \bigoplus_{\alpha\in\S}\;L_\a}\:,
\\
{\displaystyle
\S:=\{\,\a\in R_+\mid\a(H_\l)=2\,\}}\:.
\lb{ns3}
\end{array}
\renewcommand{\arraystretch}{1}
\ee
$R_+$ denotes the set of positive roots in $R$ (relative to the basis
S). In the generic case $\S$ turns out to be a basis of a root system
contained in $R$. (This is proven in Appendix A of Ref.\ \cite{OB2}.)

The $LG$-valued functions on the orbit space determine part of the
connection on $P(M,G)$. Before giving a parametrization of the YM
fields in a convenient gauge, we fix our conventions in parametrizing
the Lorentz metric $g$ on $M$ which is, of course, assumed to be
invariant under the induced SU$(2)$ action. We use standard
Schwarzschild-like coordinates and set
\be
g= -NS^2 dt^2 + N^{-1} dr^2 + r^2({d\vartheta}^2 +
\sin^2\!\vartheta\,{d\varphi}^2)\:,
\lb{ns5}
\ee
where the metric functions $N$ and $S$ depend only on $r$ and $t$. We
use also the usual mass fraction $m(r,t)$, defined by $N=:1-2m/r$.

In Ref.\ \cite{OB1} it is shown that there exists always a (local)
gauge such that the gauge potential $A$ takes the form
\be
A=\tilde{A}+\hat{A}\:,
\lb{ns6}
\ee
with
\be
\hat{A}={\Lambda_2
\,d\vartheta}+{(\Lambda_3\cos\vartheta-\Lambda_1\sin\vartheta)\,d\varphi}
\lb{ns7}
\ee
and
\be
\tilde{A}=NS{\cal A}\,dt+{\cal B}\,dr\:,
\lb{ns8}
\ee
where ${\cal A}$ and ${\cal B}$ commute with $H_\l$ (i.e. with $\L_3$).
If $H_\l$ is generic its centralizer is the infinitesimal torus $LT$.
Hence, ${\cal A}$ and ${\cal B}$ are $LT$-valued and $\tilde{A}$ is
thus Abelian.

For the example of the gauge group SU$(2)$, $H_\l$ is an integer
multiple of $4\pi\,\tau_3$: $H_\l=4\pi k\,\tau_3$ with $k\in{\cal Z}$,
and the only solutions of (\ref{ns3}) are $\L_1=\L_2=0$,
$\L_3=k\,\tau_3$, or
\be
\L_1=w\,\tau_1+\tilde{w}\,\tau_2\:,\qquad
\L_2=\mp\tilde{w}\,\tau_1\pm w\,\tau_2\:,\qquad
\L_3=\pm\tau_3\:.
\lb{ns9}
\ee

We introduce some further notation which is used below. A suitably
normalized  $\Ad (G)$-invariant scalar product on $LG$ will be denoted
by $\ldel\,\cdot\,,\cdot\,\rdel$. We use the same symbol for the
hermetian extension to $LG_{\cal C}$ (antilinear in the first
argument), and $|\cdot|$ means the corresponding norm. Note that the
original $\Ad(G)$-invariance extends on $LG_\bcal{C}$ to
\be
\sprod{X}{\lbrak{Z}{Y}}+\sprod{\,\lbrak{c(Z)}{X}}{Y}=0\:,
\lb{ns10}
\ee
where $c$ is the conjugation in $LG_\bcal{C}$.

If we insert the parametrizations (\ref{ns5})-(\ref{ns8}) into the EYM
equations, we obtain a system of partial differential equations for the
metric functions $N$, $S$ and the YM amplitudes $\L_\pm$, $\cala$,
$\calb$. For our instability proof it will suffice to write them in the
temporal gauge $\cala=0$. Specializing the results of \cite{OB1} (and
using slightly different notation) they read as follows:

The Einstein equations give two constraint equations for the $r$
derivative (denoted by a dash) and the $t$ derivative (denoted by a
dot) of $m$
\be
m'= {\kappa\over 2}
\Bigl\{
NG+r^2 p_\theta
\Bigr\}\:,
\qquad
\dot m={\kappa\over 2}NH\:,
\lb{ns11}
\ee
($\kappa:=8\pi G$), and the $(rr)$-equation reduces to
\be
{S'\over S}={\kappa\over r}G\:,
\lb{ns12}
\ee
where
\addtolength{\jot}{5pt}
\begin{eqnarray}
G &=& \frac{1}{2}
\biggl\{\,
(NS){\vphantom{|\L_+|}}^{-2}\,|\dot\L_+|{\vphantom{|\L_+|}}^2
	  +|\,\L_{+}'+[\bcal{B},\L_+]\,|{\vphantom{|\L_+|}}^2
\biggr\}\:,\lb{ns13} \\
H &=& \mbox{Re}\,
\bigl\ldel\,
\dot\L_{+}\,,\,\L_{+}'+[\bcal{B},\L_+]\,
\bigr\rdel\:,\lb{ns14}\\
p_\theta &=& {1\over 2r^{4}}
\Bigl\{
|\check {\cal F}|{\vphantom{|{\cal F}}}^2
 +|\hat {\cal F}|{\vphantom{|{\cal F}}}^2
\Bigr\}\lb{ns15}
\end{eqnarray}
\addtolength{\jot}{-5pt}
with
\be
\check{\cal F} = \frac{r^2}{S}\,\dot\bcal{B}\:,
\qquad
\hat  {\cal F} = \frac{i}{2}\,[\L_+,\L_-]-\L_3\:.
\lb{ns16}
\ee

The YM equations decompose into

\vspace{.5ex}
\be
\frac{2}{NS}
\biggl(
\frac{r^2}{S}\,\dot\bcal{B}
\biggr)\!{\vphantom{\biggr)}}^{\textstyle\cdot}
+ \Bigl[\,\L_+\,,\,\L_-'+[\bcal{B},\L_-]\,\Bigr]
+ \Bigl[\,\L_-\,,\,\L_+'+[\bcal{B},\L_+]\,\Bigr]=0\:,
\lb{ns17}
\ee

\vspace{-2.25ex}
\begin{eqnarray}
\frac{1}{S}
\biggl(
\frac{1}{NS}\,\dot\L_+
\biggr)\!{\vphantom{\biggr)}}^{\textstyle\cdot}
\!\!\!&-&\!\!\!
\frac{1}{S}
\biggl(NS\Bigl\{\,
\L_+'+[\bcal{B},\L_+]
\,\Bigr\}\biggr)'
\nonumber\\
\!\!\!&-&\!\!\!
N
\biggl\{\,
[\bcal{B},\L_+']+\Bigl[\,\bcal{B}\,,\,[\bcal{B},\L_+]\,\Bigr]
\,\biggr\}
+\frac{i}{r^2}\,[\hat {\cal F},\L_+]=0\:,
\lb{ns18}
\end{eqnarray}

\vspace{-1.5ex}
\be
2
\biggl(
\frac{r^2}{S}\,\dot\bcal{B}
{\biggr)}'
+ 2\frac{r^2}{S}[\bcal{B},\dot\bcal{B}\,]
+\frac{1}{NS}
\biggl\{\,
[\L_+,\dot\L_-]+[\L_-,\dot\L_+]
\,\biggr\}
=0\:.
\lb{ns19}
\ee

\vspace{3ex}\noindent
The last equation is the Gauss constraint. For the generic case the
term proportional to $[\bcal{B},\dot\bcal{B}\,]$ in (\ref{ns19})
vanishes.

For static solutions all time derivatives disappear and the basic
equations simplify considerably. (For the Bartnik-McKinnon solutions
$\L$ is of the form (\ref{ns9}) with $\tilde w=0$, $\L_3=\tau_3$ and
$\tilde A=0$ in (\ref{ns6}).)
\section{Perturbation equations}
In this section we study time-dependent perturbations of a given
static, asymptotically flat solution of the coupled EYM equations
(\ref{ns11}), (\ref{ns12}), (\ref{ns17})-(\ref{ns19}). Regular
solutions are of purely magnetic type ($\cala=0$ in (\ref{ns6})) with
vanishing YM charge. Unfortunately, this is not yet rigorously proven
under satisfactory weak fall-off conditions, but there is strong
evidence for this (see \cite{OB2,kunzle2} for partial results.) The
perturbation equations which we shall derive hold also for black holes,
if these are assumed to be of purely magnetic type.

{}From now on the symbols $\L_\pm$, $N$, $S$, etc. refer to the
equilibrium solution and time-dependent perturbations are denoted by
$\d\L_\pm,\d\bcal{B}$, etc.. All basic equations are linearized about
the equilibrium solution.

First, we linearize the right hand sides of the Einstein equations
(\ref{ns11}) and (\ref{ns12}). Since $\calb$ and $\dot\L_\pm$ vanish
for the equilibrium solution, the first order variation of the source
$G$ is
\be
\d G=\Re\sprod{\L'_+}{\d\L'_+}
   -\Re\sprod{\L'_+}{\lbrak{\L_+}{\d\calb\,}}\:.
\ee
Here, the last term vanishes, because the properties of the scalar
product mentioned earlier (notably (\ref{ns10})) imply that
\be
-2\Re\sprod{\L'_+}{\lbrak{\L_+}{\d\calb\,}}
=
\sprod{\,\lbrak{\L_+}{\L'_-}+\lbrak{\L_-}{\L'_+}}{\d\calb}\:,
\ee
and the YM equation (\ref{ns17}) for the equilibrium solution shows
that
\be
\lbrak{\L_+}{\L'_-}+\lbrak{\L_-}{\L'_+}=0\:.
\lb{ns20}
\ee
Thus
\be
\d G=\Re\sprod{\L'_+}{\d\L'_+}\:.
\lb{ns21}
\ee
The only first order variation for $p_\theta$ comes from $\d |\hat
{\cal F}|{\vphantom{|{\cal F}}}^2=2\sprod{\hat{\cal F}}{\d\hat{\cal
F}}$, with (see eq.\ (\ref{ns16}))
\be
\d\hat{\cal F} = \frac{i}{2}\,[\L_+,\d\L_-]
		-\frac{i}{2}\,[\L_-,\d\L_+]\:,
\lb{ns22}
\ee
giving
\be
\d p_\theta = {1\over r^{4}}\Re\sprod{i\,\lbrac{\hat{\cal
F}}{\L_+}}{\d\L_+}\:.
\lb{ns23}
\ee

Now we can work out the variation of the first Einstein equation in
(\ref{ns11}). With (\ref{ns21}), (\ref{ns23}) and (\ref{ns12}) for the
equilibrium solution, we find
\be
\d m'=
-\,{S'\over S}\,\d m
+{\kappa\over 2}\,
\biggl\{
   N\Re\sprod{\L'_+}{\d\L'_+}
+ \Re\sprod{\,{i\over r^{2}}\lbrac{\hat{\cal F}}{\L_+}}{\d\L_+}
\biggr\}\:.
\ee
For the commutator in the last term we use the unperturbed YM equation
(\ref{ns18}), i.e.
\be
\frac{i}{r^2}\,[\hat {\cal F},\L_+]=
N\frac{S'}{S}\L'_+ + N'\L'_+ + N\L''_+\:,
\lb{ns24}
\ee
whence
\be
\d m'=
-\,{S'\over S}\,\d m +
{S'\over S}\biggl\{\,
{\kappa\over 2} N\Re\sprod{\L'_+}{\d\L_+}\,\biggr\}
+
\biggl\{\,
{\kappa\over 2} N\Re\sprod{\L'_+}{\d\L_+}
\,\biggr\}'
\ee
or
\be
(\d m\,S)'=
\biggl\{\,
{\kappa\over 2} NS\Re\sprod{\L'_+}{\d\L_+}
\,\biggr\}'\:.
\ee
Therefore, $\d m$ must be of the form
\be
\d m=
{\kappa\over 2} N\Re\sprod{\L'_+}{\d\L_+}+{f(t)\over S}\:,
\lb{ns25}
\ee
where $f(t)$ is a function of $t$ alone. This function is determined by
considering the variation of the second Einstein equation in
(\ref{ns11}), which reads
\be
\d \dot{m}=
{\kappa\over 2} N\Re\sprod{\L'_+}{\d\dot{\L}_+}\:.
\ee
Thus we have also
\be
\d m=
{\kappa\over 2} N\Re\sprod{\L'_+}{\d\L_+}+g(r)\:,
\lb{ns26}
\ee
with a function $g(r)$ of $r$ alone. By comparing (\ref{ns25}) and
(\ref{ns26}) we arrive at the remarkably simple result
\be
\d m=
{\kappa\over 2} N\Re\sprod{\L'_+}{\d\L_+}\:,
\lb{ns27}
\ee
which generalizes an observation already made in \cite{NS2}.

The variation of the Einstein equation (\ref{ns12}) is immediately
obtained with (\ref{ns21})
\be
\d
\biggl(
{S' \over S}
\biggr)
={\kappa\over r} N\Re\sprod{\L'_+}{\d\L'_+}\:.
\lb{ns28}
\ee

Before also linerizing the YM equations we introduce a decomposition of
$\L_+$ and $\d\L_+$ into ``real" and ``imaginary" parts. For this we
introduce (for $\a \in \S$) a basis ${e_\a}$ of the root spaces $L_\a$
in the direct sum (\ref{ns3}) with respect to which we expand the
unperturbed $\L_+$ as well as its perturbation $\d\L_+$,
\be
  \L_+=\sum_{\a\in\S}\,   w^\a\, e_\a\:,\qquad
\d\L_+=\sum_{\a\in\S}\,\d w^\a\, e_\a\:.
\lb{ns29}
\ee
Then we have
\be
\d\L_\pm=\d X_\pm \pm i\d Y_\pm
\lb{ns30}
\ee
with
\be
\d X_+=\sum_{\a\in\S}\Re(\d w^\a)\, e_\a\:,\qquad
\d Y_+=\sum_{\a\in\S}\Im(\d w^\a)\, e_\a
\lb{ns31}
\ee
and the corresponding expansion for $\d X_-$ and $\d Y_-$ with $e_\a$
replaced by  $c(e_\a)\in L_{-\a}$, because $\d\L_-=c(\d\L_+)$ and thus
\be
\d X_-=c(\d X_+)\:,\qquad \d Y_-=c(\d Y_+)\:.
\lb{ns32}
\ee
We shall call $\d X_\pm$,  $\d Y_\pm$ the real and imaginary parts of
the perturbations  $\d \L_\pm$. It was shown in \cite{OB2} that the
unperturbed  $\L_+$ can be chosen to have only a {\em\/real\/} part.

This decomposition will lead to a significant decoupling of the
perturbation equations. Note in particular, that the variations $\d m$
and $\d p_\theta$ in (\ref{ns23}) and (\ref{ns27}) depend only on the
real part $\d X_+$:
\addtolength{\jot}{5pt}
\begin{eqnarray}
\d m       &=&
{\kappa\over 2} N\sprod{\L'_+}{\d X_+}\:,
\lb{ns33}\\
\d p_\theta&=&
{1\over r^{4}}\sprod{i\,\lbrac{\hat{\cal F}}{\L_+}}{\d X_+}\:.
\lb{ns34}
\end{eqnarray}
\addtolength{\jot}{-5pt}

We consider now the first variation of the YM equation (\ref{ns18}).
Its decomposition into real and imaginary parts yields
\addtolength{\jot}{5pt}
\begin{eqnarray}
-\,\frac{1}{NS^2}\,\d \ddot{X}_+&=&
-N\d X''_+
\,-\,\frac{(NS)'}{S}\d X'_+
\,-\,{i\over r^{2}}\lbrac{\L_+}        {\d\hat{\cal F}\,}
\,+\,{i\over r^{2}}\lbrac{\hat{\cal F}}{\d\L_+}
\nonumber\\
&&
\mbox{}\,-\,\d N\L''_+
\,-\,\d
\biggl(
{(NS)' \over S}
\biggr)
\L'_+
\lb{ns35}
\end{eqnarray}
and
\begin{eqnarray}
-\,\frac{1}{NS^2}\,\d \ddot{Y}_+&=&
-N
\biggl\{\,
\d Y''_+
\,+\,i\,\lbrak{\L_+}{\d\bcal{B}}'
\,+\, i\,\lbrak{\L_+'}{\d\bcal{B}}
\,\biggr\}
\nonumber\\
&&
\mbox{}\,-\,{(NS)' \over S}
\biggl\{\,
\d Y'_+
\,+\,i\,\lbrak{\L_+}{\d\bcal{B}}
\,\biggr\}
\,+\,\frac{i}{r^2}\,[\hat {\cal F},\d Y_+]\:.
\lb{ns36}
\end{eqnarray}
\addtolength{\jot}{-5pt}
The third term on the right hand side of (\ref{ns35}) is indeed real
and can be written, using (\ref{ns21}), as
\be
 \frac{i}{r^2}\,[\L_+,\d\hat {\cal F}\,]=
\frac{1}{r^2}\,\ad(\L_+)\,\ad(\L_-)\,\d X_+\:.
\lb{ns37}
\ee
Equation (\ref{ns35}) can be simplified further. {}From (\ref{ns33}) and
the equilibrium equation (\ref{ns24}) we deduce
\addtolength{\jot}{5pt}
\begin{eqnarray*}
-\,\d N\L''_+&=&\frac{2}{r}\,\d m\,\L''_+\\
	     &=&\kappa N\Re\sprod{\L'_+}{\d X_+}\:\L''_+\\
	     &=&\kappa  \Re\sprod{\L'_+}{\d X_+}\:
		\biggl\{\,
		-{(NS)' \over S}\L_+
		\,+\,\frac{i}{r^2}\,[\hat {\cal F},\L_+]
		\,\biggr\}\:,
\end{eqnarray*}
\addtolength{\jot}{-5pt}
and the Einstein equations (\ref{ns11}), (\ref{ns12}) give
\be
-\,\d\,{(NS)' \over S}=-\frac{2}{r^2}\d m + \kappa r\,\d p_\theta\:.
\ee
If we use here  (\ref{ns33}) and (\ref{ns34}) we see that the last two
terms in  (\ref{ns35}) can be expressed as follows:
\addtolength{\jot}{5pt}
\begin{eqnarray}
&&\quad\llap{${}-\,\d N\L''_+$} \,-\, \d\,{(NS)' \over S}
=\frac{1}{NS^2}
\Biggl\{\;
-(\,p_\ast\L_+)\:\frac{\kappa}{r}
\biggl\{
\frac{(NS)'}{NS}+\frac{1}{r}
\biggr\}
\sprod{\,p_\ast\L_+}{\d X_+}\nonumber\\
&&\qquad\qquad
\llap{${}+(\,p_\ast\L_+)$}
\:\frac{\kappa S}{r^3}
\sprod{\,\lbrak{\hat\bcal{F}}{\L_+}}{\d X_+}
+\:\lbrak{\hat\bcal{F}}{\L_+}
\:\frac{\kappa S}{r^3}
\sprod{\,p_\ast\L_+}{\d X_+}
\:\:\Biggr\}\:\:,
\lb{ns38}
\end{eqnarray}
\addtolength{\jot}{-5pt}
where we have introduced the differential operator
\be
p_\ast=-iNS\frac{\partial}{\partial r}\;.
\lb{ns39}
\ee
Inserting these expressions into (\ref{ns35}) gives the following
pulsation equation for the real amplitude $\d X_+$ of the YM field
\be
\d \ddot X_+ +\,U_{XX}\,\d X_+=0\:,
\lb{ns40}
\ee
where the operator $U_{XX}$ is given by
\addtolength{\jot}{8pt}
\begin{eqnarray}
U_{XX}&=&
{p_\ast}^2\,+\,\frac{NS^2}{r^2}\ad(i\hat\bcal{F})
\,-\,\frac{1}{Nr^2}\,\ad(NS\L_+)\,\ad(NS\L_-)\nonumber\\
&&{}\:-\:(\,p_\ast\L_+)\:\frac{\kappa}{r}
\biggl\{
\frac{(NS)'}{NS}+\frac{1}{r}
\biggr\}
\sprod{\,p_\ast\L_+}{\cdot\,}\nonumber\\
&&{}\,+\:(\,p_\ast\L_+)
\:\frac{\kappa S}{r^3}
\sprod{\,\lbrak{\hat\bcal{F}}{\L_+}}{\cdot\,}
\:+\:\lbrak{\hat\bcal{F}}{\L_+}
\:\frac{\kappa S}{r^3}
\sprod{\,p_\ast\L_+}{\cdot\,}\lb{ns41}\:.
\end{eqnarray}
\addtolength{\jot}{-8pt}

It is remarkable that the perturbations $\d Y_\pm$ and $\d\calb$ do not
appear in (\ref{ns40}) and that the back reaction of gravitation on $\d
X_+$ can be described by an effective potential (last three terms in
(\ref{ns41})).

Equation (\ref{ns36}) can easily be brought into the form
\be
\d \ddot Y_+ +\,U_{YY}\,\d Y_+ +\,U_{Y\calb}\,\d \calb=0\:,
\lb{ns42}
\ee
where
\addtolength{\jot}{5pt}
\begin{eqnarray}
U_{YY}
&=&{p_\ast}^2\,+\,\frac{NS^2}{r^2}\ad(i\hat\bcal{F})\:,\lb{ns43}\\
U_{Y\calb}&=& p_\ast\,\ad(NS\L_+)\,+\,\ad(NS\,p_\ast\L_+)\:.\lb{ns44}
\end{eqnarray}
\addtolength{\jot}{-5pt}
We have thus achieved a partial decoupling, because neither $\d X_+$,
nor the metric perturbations appear in (\ref{ns42}).

We proceed with the linearization of the YM equation (\ref{ns17}). The
variation of the last two terms is
\be
-\Bigl[\,\L_+\,,\,[\L_-,\d\bcal{B}]\,\Bigr]
\,+\,
[\L_+,\d\L'_-]\,-\,[\L'_-,\d\L_+]
\,+\,
\mbox{conjugate}\:,
\ee
which leads (with $\d\L_\pm=\d X_\pm \pm i\d Y_\pm$) to
\addtolength{\jot}{8pt}
\begin{eqnarray*}
&&-\,\biggl\{
\Bigl[\,\L_+\,,\,[\L_-,\d\bcal{B}]\,\Bigr]
+i\,[\L_+,\d Y'_-]+i\,[\L'_-,\d Y_+]
\:\biggr\}\\
&&
\qquad\hphantom{-\Bigl[\,\L_+\,,\,[\L_-,\d\bcal{B}]\,\Bigr]}
{}\,+\,\biggl\{\,
[\L_+,\d X'_-]-[\L'_-,\d X_+]\,\biggr\}
\,+\,
\mbox{conjugate}\:.
\end{eqnarray*}
\addtolength{\jot}{-8pt}
Here, the terms in the first curly bracket are in $LT$, while those in
the second are in $iLT$. The latter are compensated by their conjugates
and we find
\be
Nr^2\d \ddot \calb_+ +\,U_{\calb\calb}\,\d \calb_+ +\,U_{\calb Y}\,\d
Y_+=0\:,
\lb{ns45}
\ee
with
\begin{eqnarray}
U_{\calb\calb}&=& -\:\ad(NS\L_+)\,\ad(NS\L_-)\:,\lb{ns46}\\
U_{\calb Y}&=& -\:\ad(NS\L_-)\,p_\ast\,+\,\ad(NS\,p_\ast\L_-)\lb{ns47}
\end{eqnarray}

At this point we collect the results obtained so far as follows: Let
\be
\renewcommand{\arraystretch}{1.5}
\d\Psi =\left( \begin{array}{c}
\d Y_+ \\
\d\bcal{B}
\end{array} \right)\:, \qquad\qquad
T=\left( \begin{array}{cc}
1 & 0 \\
0 & Nr^2
\end{array} \right)\:,
\lb{ns48}
\renewcommand{\arraystretch}{1}
\ee
then (\ref{ns42}) and (\ref{ns45}) can be written as a $2\times 2$
matrix equation
\be
T\d \ddot\Psi + U\d\Psi=0\:,
\lb{ns49}
\ee
with
\be
\renewcommand{\arraystretch}{1.5}
U=\left( \begin{array}{cc}
     U_{YY} & U_{Y\calb} \\
U_{\calb Y} & U_{\calb\calb}
\end{array} \right)\:.
\lb{ns50}
\renewcommand{\arraystretch}{1}
\ee
The operators in this matrix are given in eqs. (\ref{ns43}),
(\ref{ns44}), (\ref{ns46}) and (\ref{ns47}).

The perturbation equations (\ref{ns40}) and (\ref{ns49}) do not include
the Gauss constraint (\ref{ns19}), whose linearization is easily found
to be
\be
\frac{\partial}{\partial t}
\biggl\{\,
p_\ast\biggl(\frac{r^2}{S}\d\calb\biggr)-\lbrak{\L_+}{\d Y_-}
\,\biggl\}
=0\:.
\lb{ns51}
\ee
The role of this constraint will be discussed later.

In concluding this section we emphasize once more, that the
perturbation equations hold also for black holes, if these are assumed
to be of purely magnetic type. In our further discussion we will,
however, consider only perturbations of uncharged regular solutions.
\section{The eigenvalue problem}
It is natural to introduce the following scalar product for
$LG_\bcal{C}$-valued functions on $\bcal{R}_+$:
\be
\braket{\phi}{\psi}=\int_0^\infty\sprod{\phi}{\!\psi}\;\,\frac{dr}{NS}\:,
\lb{ns52}
\ee
because the operators $U_{XX}$, $U$ and $T$ are symmetric with respect
to this scalar product on a dense domain of $\bmrm{L}^2$-functions.
This can easily be seen by using
\be
\braket{\phi}{p_\ast\psi}=\braket{p_\ast\phi}{\psi}
\lb{ns53}
\ee
for smooth functions which vanish at the origin, and
\be
\braket{\phi}{\ad (Z)\psi}=-\braket{\ad (c(Z))\phi}{\psi}
\lb{ns54}
\ee
for all $LG_\bcal{C}$-valued functions $\phi$, $\psi$, $Z$ in
$\bmrm{L}^2$ (see (\ref{ns10})).

We specialize now to harmonic perturbations
\be
\d X_+(r,t)=\xi(r)\mbox{e}^{-i\omega t}\:,\qquad
\d \Psi(r,t)=\d\Phi(r)\mbox{e}^{-i\omega t}\:,
\lb{ns55}
\ee
whose frequencies satisfy the eigenvalue equations
\be
U_{XX}\xi=\omega^2\xi\:,
\lb{ns56}
\ee
and
\be
U\d\Phi=\omega^2\,T\,\d\Phi\:.
\lb{ns57}
\ee

It should be remarked at this point that (\ref{ns41}) and (\ref{ns56})
reduce for the Bartnik-McKinnon solution \cite{bartnik} to the
eigenvalue problem derived in Ref.\ \cite{NS2}, where it was shown that
this has exactly one unstable mode. (A similar instability for the
``colored" black hole was found in \cite{NS3}.)

Let us now turn to the role of the linearized Gauss constraint
(\ref{ns51}) in conjunction with the eigenvalue problem (\ref{ns57}).
We show first that a variation $\d\Phi$ is orthogonal with respect to
the scalar product defined by $T$,
\be
\braket{\,\cdot\,}{\,\cdot\,}_T=\bra{\,\cdot\,}\:T\,\ket{\,\cdot\,}\:,
\lb{ns58}
\ee
to all gauge variations
\be
\renewcommand{\arraystretch}{1.5}
\d\Phi_{\bmrm{gauge}} =\left( \begin{array}{c}
i\,\lbrac{\chi}{\L_+}\\
\chi'
\end{array} \right)\:,
\lb{ns59}
\renewcommand{\arraystretch}{1}
\ee
if and only if the curly bracket in (\ref{ns51}) vanishes. Note that
these gauge variations arise if (\ref{ns6}) is subjected to the gauge
transformation $g=\exp(\epsilon\chi)$, because (\ref{ns7}) and
(\ref{ns8}) show that this induces the infinitesimal transformation
\be
\L_+\rightarrow\L_+-\lbrac{\chi}{\L_+}\:,\qquad
\calb \rightarrow \calb+\chi'\:.
\ee
To prove the statement just made, we compute
\addtolength{\jot}{8pt}
\begin{eqnarray}
\bra{\d\Phi}\,T\:\ket{\d\Phi_{\bmrm{gauge}}}&=&
\vphantom{+}\int_0^\infty
\:\Bigl\{\,
 \sprod{\d Y_+}{-i\,\lbrac{\L_+}{\chi}}
+\sprod{\d\calb}{Nr^2\chi'}
\,\Bigr\}\,\frac{dr}{NS}\:,\nonumber\\
&=&
-\int_0^\infty
\:\Bigsprod
{\,\biggl(\frac{r^2}{S}\d\calb\biggr)'+\frac{i}{NS}\lbrak{\L_-}{\d
Y_+}\,}
{\chi}
\,dr\:.
\lb{ns60}
\end{eqnarray}
\addtolength{\jot}{-8pt}
Here we have used (\ref{ns10}) and made a partial integration, dropping
a boundary term. This is allowed if $\chi$ is regular at the origin and
vanishes sufficiently fast at infinity. Since $\chi$ is otherwise
arbitrary and $i\,\lbrac{\L_-}{\d Y_+}=-i\,\lbrac{\L_+}{\d Y_-}$,
eq.\ (\ref{ns60}) implies our claim.

In the next section we will show that the eigenvalue equation
(\ref{ns57}) has at least one mode $\d\Phi$ with $\omega^2 < 0$. Such a
mode is orthogonal with respect to the scalar product (\ref{ns58}) to
any zero mode of (\ref{ns57}) and thus in particular to
$\d\Phi_{\bmrm{gauge}}$ in (\ref{ns59}). (This follows since different
eigenmodes of (\ref{ns57}) are orthogonal with respect to the scalar
product (\ref{ns58}), because $U$ and $T$ are symmetric with respect to
(\ref{ns52}).) Hence, we can conclude that the Gauss constraint is
automatically satisfied.

{}From (\ref{ns57}) we obtain
\be
{\omega}^2 =\: \frac{\bra{\d\Phi}\:U\,\ket{\d\Phi}}
		      {\bra{\d\Phi}\:T\,\ket{\d\Phi}}\:,
\lb{ns61}
\ee
and for the frequence $\omega_0$ of the fundamental mode we have the
minimum principle
\be
{\omega_0}^2 =\Min_{\d\Phi}\: \frac{\bra{\d\Phi}\:U\,\ket{\d\Phi}}
			      {\bra{\d\Phi}\:T\,\ket{\d\Phi}}\:.
\lb{ns62}
\ee

We do not discuss here the precise mathematical nature of the
eigenvalue problem (domains of definition, essential selfadjointness,
etc.), because the functional analytic aspects are very similar to
other well-studied eigenvalue problems.
\section{Instability of generic EYM solitons}
We are now ready to establish the main point of this paper:

For a given regular solution with $\L_+=\sum_{\a\in\S}\, w^\a e_\a$ we
shall construct a one-parameter family of field configurations
$\L_{(\chi)+}$, ${\calb}_{(\chi)}$ such that the variational
expressions (\ref{ns62}) for
$\d\L_\pm=(d\L_{(\chi)\pm}/d\chi)_{\chi=0}$,
$\d\calb=(d{\calb}_{(\chi)}/d\chi)_{\chi=0}$ is {\em\/ negative \/}.
This family is chosen of the following form:
\bea
\L_{(\chi)+}&=&\Ad(\exp(-\chi Z))
\Bigl\{\,
\L_+\cos(\chi)+i\L_+(\infty)\sin(\chi)
\,\Bigr\} \lb{ns63}\:,\\
\calb_{(\chi)}&=&\chi Z'\:,
\lb{ns64}
\eea
where $Z$ is an $LT$-valued function of $r$ with the properties
\be
\lim_{r\to 0,\infty}\:\lbrak{Z}{\L_+}=i\L_+(\infty)\:,\qquad
\supp Z'\subseteq[\,1-\epsilon\,,1+\epsilon\,]
\lb{ns65}
\ee
for an $\epsilon >0$. The existence of such a function can be seen as
follows: Let $\{h_\a\}_{\a\in\S}$ be the dual basis of $2\pi\S$ and put
\be
Z=\sum_{\a\in\S}\, Z^\a h_\a\:,\qquad
Z^\a  =  \left \{
\begin{array}{ll}
w^\a (\infty)/w^\a (0)   & \mbox{ for }r<1-\epsilon\:,\\
1                        & \mbox{ for }r>1+\epsilon\:.
\end{array} \right.
\lb{ns66}
\ee
It is easy to verify that both conditions in (\ref{ns65}) are
satisfied. (In Appendix A of ref.\ \cite{OB2} we have shown that $w^\a
(0)\neq 0 \mbox{ for all } \a\in\S$.)

We note a few properties of the family (\ref{ns63}), (\ref{ns64}). The
equilibrium solution is clearly obtained for $\chi=0$. Applying a gauge
transformation with $g=\exp(-\chi Z)$ we obtain
\be
\L_{(\chi)+}\to\L_+\cos(\chi)+i\L_+(\infty)\sin(\chi)\:,\qquad
\calb_{(\chi)}\to 0\:.
\lb{ns67}
\ee
The first variations of (\ref{ns63}) and (\ref{ns64}) are
\be
\d\L_+=-\lbrak{Z}{\L_+}+i\L_+(\infty)\:,\qquad
\d\calb=Z'\:,
\lb{ns68}
\ee
and these satisfy by construction the desired boundary conditions
\be
\lim_{r\to 0,\infty}\d\L_+=0\:,\qquad
\lim_{r\to 0,\infty}\d\calb=0\:.
\lb{ns69}
\ee
($\d\calb$ vanishes even outside $[\,1-\epsilon\,,1+\epsilon\,]$.)
Finally, $\d\L_+$ has only an imaginary component
\be
\d Y_+=-i\d\L_+=i\,\lbrak{Z}{\L_+}+\L_+(\infty)
\lb{ns70}
\ee
and thus by (\ref{ns62})
\be
{\omega_0}^2 \leq\: \frac{\bra{\d\Phi}\:U\,\ket{\d\Phi}}
		    {\bra{\d\Phi}\:T\,\ket{\d\Phi}}\:.
\lb{ns71}
\ee
with $\d\Phi=(\d Y_+,\d\calb)$ given by (\ref{ns70}) and the second
eq.\ in (\ref{ns68}).

This judicious choice of trial functions fulfills our goal: The
denominator in (\ref{ns71}) is finite and the numerator turns out to be
strictly negative!

The first of these two points is simple. Since $\d \calb$ in
(\ref{ns68}) has compact support, we have to check only whether
\be
\int_0^\infty
\,\sprod{\d Y_+}{\d Y_+}\,\frac{dr}{NS}\:<\:\infty\:.
\ee
By construction, $i\d Y_+=\L_+(r)-\L_+(\infty)$ for $r>1+\epsilon$.
Since $N$ and $S$ both approach $1$ at infinity, the integral is finite
if $\L_+(r)-\L_+(\infty)$ is assumed to converge to zero faster than
$r^{-1/2}$.

The calculation of the numerator in (\ref{ns71}) is somewhat tedious.
Considerable simplifications occur by separating a gauge mode in
$\d\Phi$:
\be
\renewcommand{\arraystretch}{1.5}
\d\Phi=
\d\Phi_{\bmrm{gauge}} +
\left(
\begin{array}{c}
\L_+(\infty)\\
0
\end{array}
\right)\:,\qquad
\d\Phi_{\bmrm{gauge}} =
\left(
\begin{array}{c}
i\,\lbrac{Z}{\L_+}\\
Z'
\end{array}
\right)\:.
\lb{ns72}
\renewcommand{\arraystretch}{1}
\ee
Clearly $U\d\Phi_{\bmrm{gauge}}=0$, and thus (\ref{ns43}) and
(\ref{ns47}) give
\be
\renewcommand{\arraystretch}{1.5}
U\d\Phi\,=
\left(
\begin{array}{c}
U_{YY}     \L_+(\infty)\\
U_{\calb Y}\L_+(\infty)
\end{array}
\right)
\,=
\frac{NS^2}{r^2}
\left(
\begin{array}{c}
i\,\lbrac{\hat{\bcal{F}}}{\L_+(\infty)}\\
-iNr^2\lbrac{\L_-'}{\L_+(\infty)}
\end{array}
\right)\:.
\lb{ns73}
\renewcommand{\arraystretch}{1}
\ee
{}From (\ref{ns72}) and (\ref{ns73}) we obtain
\addtolength{\jot}{10pt}
\begin{eqnarray}
\bra{\d\Phi}\:U\,\ket{\d\Phi}
&=&
\int
\:\sprod{\,i\,\lbrac{Z}{\L_+}}{i\,\lbrac{\hat{\bcal{F}}}{\L_+(\infty)}\,}
\:\frac{S}{r^2}\:\,dr
\nonumber\\
&+&
\int
\:\sprod{\L_+(\infty)}{i\,\lbrac{\hat{\bcal{F}}}{\L_+(\infty)}\,}
\:\frac{S}{r^2}\:\,dr
\nonumber\\
&+&
\int
\:\sprod{Z'}{-i\,\lbrac{\L_-'}{\L_+(\infty)}\,}
\:NS\:\,dr\:\:.
\lb{ns74}
\end{eqnarray}
\addtolength{\jot}{-10pt}
Let us denote the integrands of the three terms by $J_1$, $J_2$ and
$J_3$. We find immediately
\addtolength{\jot}{5pt}
\begin{eqnarray}
J_1
&=&\frac{S}{r^2}\,
\sprod{\hat{\bcal{F}}}
{[\,\L_+(\infty)\,,\,\lbrac{Z}{\L_-}\,]\,}\:,
\lb{ns75}\\
J_2
&=&2\,\frac{S}{r^2}
\sprod{\hat{\bcal{F}}}{\L_3}\:.
\lb{ns76}
\end{eqnarray}
\addtolength{\jot}{-5pt}
In the second equation we have used (\ref{ns16}) and the vanishing of
the YM charge, implying that $\lim_{r\rightarrow\infty}\L(r)$ is a
homomorphism from $L$SU(2) to $LG$, whence
\be
i\,\lbrac{\L_+(\infty)}{\L_-(\infty)}=2\L_3\:.
\lb{ns77}
\ee

In a next step we show that the first and the last term in (\ref{ns74})
compensate each other. For this we rewrite the third term, performing a
partial integration and making use of the equilibrium equation
(\ref{ns24}), as follows
\addtolength{\jot}{10pt}
\begin{eqnarray}
\int J_3 \:dr
&=&
i\int\:
\sprod{NS\L'_+}{{\lbrac{Z}{\L_+(\infty)}\,}'}
\:\,dr
\nonumber\\
&=&
i\,\sprod{NS\L'_+}{\lbrac{Z}{\L_+(\infty)}}\Bigm|_0^\infty\nonumber\\
&&\qquad-\:
\int\:
\sprod{\hat{\bcal{F}}}
{[\,\L_-\,,\,\lbrac{Z}{\L_+(\infty\,)}\,]}
\:\frac{S}{r^2}\:\,dr\:\:.
\end{eqnarray}
\addtolength{\jot}{-10pt}
The boundary term vanishes and the double commutator in the last term
is equal to $[\,\L_+(\infty)\,,\,\lbrac{Z}{\L_-}\,]$, as is seen by
using the Jacobi identity and the fact that
$\lbrac{\L_+(\infty)}{\L_-}$ is in $iLT$. Comparing this result with
(\ref{ns75}) shows that there remains indeed only the second term in
(\ref{ns74}). Thus from (\ref{ns76}) we obtain the intermediate result
\be
\bra{\d\Phi}\:U\,\ket{\d\Phi}=2\int\;\sprod{\hat{\bcal{F}}}{\L_3}\,\frac{S}{r^2}\:dr\:.
\lb{ns78}
\ee

Finally, we show that the last expression can be transformed into a
form with a definite sign:
\be
2\int\;\sprod{\hat{\bcal{F}}}{\L_3}\,\frac{S}{r^2}\:dr
=
-\int\;\Bigl\{\,
NS|\L_{+}'|{\vphantom{|\L_+|}}^2
+2\frac{S}{r^2}|\hat {\cal F}|{\vphantom{|{\cal F}}}^2
\,\Bigr\}\;dr
\:.
\lb{ns79}
\ee
In order to see this we perform a partial integration of the first term
on the right and use again the YM equation (\ref{ns76}):
\be
\int\,\sprod{\L_+'}{NS\L_+'}\;dr
= \sprod{\L_+}{NS\L_+'}\Bigm|_0^\infty
\,-\,
\int\,\sprod{\L_+}{i\,\lbrac{\hat{\cal
F}}{\L_+}\,}\,\frac{S}{r^2}\;dr\:.
\ee
The boundary term vanishes and the integral combines with the last term
in (\ref{ns79}) to the left hand side, because we have (see
(\ref{ns16}))
\be
2|\hat {\cal F}|{\vphantom{|{\cal F}}}^2
=\sprod{\hat {\cal F}}{i\,\lbrac{\L_+}{\L_-}-\L_3}
=\sprod{\L_+}{i\,\lbrac{\hat {\cal F}}{\L_+}}
-\sprod{\hat {\cal F}}{\L_3}\:.
\lb{ns80)}
\ee

All together, we have established the crucial result
\be
\bra{\d\Phi}\:U\,\ket{\d\Phi}=
-\int\;\Bigl\{\,
NS|\L_{+}'|{\vphantom{|\L_+|}}^2
+2\frac{S}{r^2}|\hat {\cal F}|{\vphantom{|{\cal F}}}^2
\,\Bigr\}\;dr\:.
\lb{ns81}
\ee
This expression is finite and strictly negative. Hence we have shown
that $\omega_0$ is indeed negative, and thus there exist unstable modes
of (\ref{ns57}). These fulfill, we recall, automatically the linearized
Gauss constraint (\ref{ns51}).

One can show that the expression (\ref{ns81}) is also equal to the
second variation of the Schwarzschild mass for the one-parameter family
(\ref{ns61}), (\ref{ns64}). (This is the way we arrived originally at
the variation (\ref{ns64})). For a systematic discussion of the
relation between variational principles for the spectra of radial
pulsations and second variations of the total mass, we refer to
\cite{OB4}.\newpage

In summary, we have proven the instability of all generic, regular
equilibrium solutions. More precisely, we have established the
following

\vskip 1ex
\noindent%
{\bf Theorem.}\hskip 1em{\em A static, sperically symmetric,
asymptotically flat, regular solution of the EYM eqs. (\ref{ns11}),
(\ref{ns12}), (\ref{ns17})-(\ref{ns19}) is {\em\/unstable\/} if the
following three conditions are satisfied:

\begin{list}{(\roman{c_one})}
	    {\usecounter{c_one}
	     \setlength{\labelwidth}{\parindent}
	     \setlength{\listparindent}{0cm}
	     \setlength{\itemindent}{0cm}
	      }

\item The solution is generic (i.e. the classifying element
$H_\lambda=-4\pi\L_3$ lies in the {\em\/open\/} fundamental Weyl
chamber).

\item The (magnetic) YM charge vanishes (i.e. $\lim_{r\to\infty}\L(r)$
is a      homomorphism from $L${\rm SU}$(2)$ to $LG$).

\item Asymptotically $\L_+(r)-\L_+(\infty)\sim r^{-\a}$ with $\a>1/2$.

\end{list}
}

\vskip 1ex
We emphasize again the strong evidence, that the assumptions of the
theorem together with condition (i) already {\em\/imply\/} condition
(ii).
Also, we would like to stress that we were able to draw this
conclusion
assuming only weak asymptotic conditions for the solitons. In
particular, the fall-off conditon (iii) of the theorem is mild and is
certainly fulfilled for the Bartnik-McKinnon solutions, as was shown
rigorously in \cite{maison}. The same is true for the solutions which
have been found numerically by H.P.\ K\"unzle for the group SU$(3)$
\cite{kunzle2}. (For both types the exponent $\alpha$ is equal to one.)
\section*{Acknowledgments}
We would like to thank Markus Heusler for discussions and comments and
to Michael Volkov for suggestive remarks on stability problems. This
work was supported by the Swiss National Science Foundation.

\end{document}